\documentclass[11pt]{revtex4}
\usepackage{hyperref}
\usepackage{pdflscape} 
 \usepackage{graphics}
 \usepackage{subfigure}
\usepackage{epsfig}
 \usepackage{amsmath,amsfonts}
 \usepackage{amssymb}
\begin{document} 
\title{Stability of  remnants of Bardeen regular black holes in presence of thermal fluctuations} 
\author{Yawar H. Khan$^{a}$}
\email{iyhkphy@gmail.com}

\author{Sudhaker Upadhyay$^{b,c,d,e}$}
\email{sudhakerupadhyay@gmail.com; sudhaker@associates.iucaa.in}

\author{Prince A. Ganai$^{a}$}
\email{princeganai@nitsri.com}
\affiliation{${}^{a}$Department of Physics, National Institute of Technology, Srinagar, Kashmir-190006, India }
\affiliation{${}^{b}$Department of Physics, K.L.S. College,  Nawada-805110, India}
\affiliation{${}^{c}$Department of Physics, Magadh University, Bodh Gaya, Bihar 824234, India}
\affiliation{${}^{d}$Inter-University Centre for Astronomy and Astrophysics (IUCAA) Pune-411007, Maharashtra, India}
\affiliation{${}^{e}$School of Physics, Damghan University, P. O. Box 3671641167, Damghan, Iran}

	\begin{abstract}
We discuss remnants of the Bardeen regular black hole motivated by using the concept of thermal fluctuations. Firstly, we derive the equilibrium values of various thermodynamic quantities like entropy, Hawking temperature, pressure, internal energy, Helmholtz free energy, and Gibbs free energy in the non-extended phase space. We then discuss Geometrothermodynamics (GTD) of Bardeen black hole to study its stability. Next we estimate the size of black hole remnant in terms of some known parameters of the black hole solution. Motivated by the fact that estimation of size, characteristics and stability of remnants of black holes could further increase our understanding of binary collisions, information loss paradox and dark energy, the black hole remnant, which gives an idea about stable mass left over after evaporation of black hole is seen to owe its presence due to thermal fluctuations. We see that the thermal fluctuations bring an overall increase in entropy curve. However, in presence of thermal fluctuations, a positive kink, which signifies a maximum increase in the value of entropy occurs at a certain value of horizon which is exactly equal to the remnant radius. We observe that the thermal fluctuations, which are characteristics of quantum gravity, lead to stable values of thermodynamic quantities near the remnant radius. In presence of thermal fluctuations we then derive various corrected thermodynamic potentials and also discuss the validity of first law of black hole thermodynamics for Bardeen black hole.  
	\end{abstract}
\maketitle
	\section{Introduction} 
	Black holes which are the most fascinating predictions of Einstein's theory of general relativity owe their presence due to the singularities in the solutions of Einstein's field equations. It is argued that global hyperbolicity in any kind of matter contracting in general relativity regime could be the main cause of these singularities \cite{1}. Further Penrose cosmic censorship gives a meaningful interpretation to these singularities by associating event horizon to every black hole singularity \cite{2}. In an attempt to overturn the argument that solutions to Einstein's field equations are always singular, Bardeen proposed a new kind of black hole called "regular black hole" \cite{3}. Some earlier attempts to replace singularity by matter backgrounds were done by Sakharov \cite{4} and Gliner \cite{5}. Later on, many more regular black hole models were then proposed \cite{6,7,8,9}. Collectively all these models came to be known as "Bardeen black holes" \cite{10,11}. Regular black holes violate some conditions of Hawking-Penrose theorems of singularity \cite{12}. The regular black holes with spherical symmetry were reported by Dymnikova \cite{13,14,15}, Bronnikov \cite{16}, and Hayward \cite{17}. Some black holes with axial symmetry could be found in \cite{18,19}. There are regular black holes with rotation given in \cite{20,21,22,23}, which in general violate weak energy condition.	
	Remarkably, Hawking found out that black holes can radiate \cite{24}. This phenomenon is now termed as Hawking radiation. In fact, black holes show characteristics of thermodynamic systems. In Ref. \cite{23a},  it was argued that the area of the black hole covered with the event horizon never decreases which finds an analogy with the entropy of any thermodynamical system. Bardeen formulated four laws of black hole mechanics \cite{25}, which gave black hole thermodynamics a significant place in black hole physics. The entropy of black holes which follows from Bekenstein's area-law \cite{26} is governed by generalized second law which assigns entropy a maximum value. Originally,  Hawking and Page \cite{27}  obtained partition function and ensemble for black holes in AdS space and naturally quantities like internal energy and temperature. The thermodynamics for regular class of black holes including the Bardeen black hole can be found in \cite{28,29,30}. In \cite{31,32,33} the corrected form of the first law of thermodynamics for regular black holes has been derived. Thermodynamics of cosmological regular black holes was done by Dymnikova et al \cite{34}.\\ 
	In the expansion of the partition function, it is found that the entropy of the black hole gets modified or corrected. These expansion terms were then justified to be there due to small thermal fluctuations around the equilibrium value of entropy. These corrections to the entropy lead to the modification of various thermodynamic equations of state. The effect of these small thermal fluctuations is by now studied in detail for various black hole systems \cite{sud,sud1,sud2,sud3,sud4}. To be more specific, thermal fluctuations in a charged AdS black hole are studied in \cite{34q}.  The effects of thermal fluctuations on the thermodynamics of the modified Hayward black hole are carried in \cite{34w}. The thermal fluctuations in Godel black hole \cite{34aa}, massive black hole in AdS space \cite{34ab}, Schwarzchild Beltrami-de Sitter black hole \cite{34ac}, dilatonic black hole \cite{34ad}  and \cite{80,81,82} are some of the examples of thermal fluctuations in black hole thermodynamics. Thermal fluctuations in a regular class of black holes, in particular Hayward black hole, are also studied  \cite{34w}. Similar studies for ds black holes in massive gravity and skyrmion black holes have been done in Ref. \cite{34aff, 34afff}. The study of effect of thermal fluctuations on thermodynamics of various black hole geometries has been recently done in detail by B. Pourhassan et al \cite{v1,v2,v3,v4,v5}. The study of higher order thermal fluctuations on thermodynamics of black holes have been studied in \cite{in1,34ab, bph, www}.

	An important aspect in the field of black hole thermodynamics is the study of remnants. The study of remnants of black holes, done in \cite{83,84} discusses the possibility of remnants as a source of dark matter. In \cite{suv}, remnants of Hayward black holes were investigated. The remnants of black holes in gravity's rainbow are calculated in \cite{xuv}. Remnants of black holes under different conditions have been calculated in Refs. \cite{90,91,92}. The study of remnants of regular black holes in presence of thermal fluctuations are still by and large unexplored, which thus is a missing link for us to explore.
	
	In this paper, we emphasize the formation of black hole remnant in Bardeen regular black hole. The radius of event horizon is given for which  the formation of stable remnant occurs. The remnant mass becomes stable and the thermodynamic quantities take the maximum possible stable values at this  horizon radius. The behaviors of Hawking temperature and mass with respect to horizon radius are discussed.  Moreover, we  calculate the leading-order corrections to the equilibrium entropy and, therefore, various thermodynamical quantities of Bardeen regular black hole described by correction parameter $\alpha$ in the entropy equation of Bardeen black hole. The possible values of $\alpha$ are $0$ and $1/2$, which describe absence and presence of correction terms, respectively. Here, we observe that the change due to correction term helps in stabilizing  Bardeen black
	hole.  Remarkably, correction term ensures the positive entropy even when black hole sinks
	towards becoming point like.  We also identify the region of stability, which is exactly the black hole remnant radius for Bardeen black hole. We then calculate effect of correction on internal energy of Bardeen regular black hole. We see that internal energy of black hole reduces in the presence of thermal fluctuations. The expression for corrected Helmholtz free energy is also obtained under the effect of thermal fluctuations. In contrast to equilibrium value, the corrected free energy of black hole becomes (asymptotically) large when horizon radius shrinks to zero. The corrected  values of pressure, enthalpy and Gibbs  free energy are also computed. Incidentally, this pressure becomes negative while size of black hole reduces. The thermal fluctuations decrease the value of enthalpy of the system. Overall the presence of thermal fluctuations  increases the stability of black holes. \\
 	The paper is presented as follows. In section \ref{2}, we review the basic thermodynamics of Bardeen regular black hole and analyze its geometrothermodynamics. We then calculate the remnant for Bardeen regular black hole in section \ref{22}. In section \ref{3}, we expand the entropy equation to get thermal fluctuation for the black hole  system. In section \ref{4}, we calculate corrected thermodynamic quantities of  Bardeen regular black hole and plot various corrected thermodynamic quantities against event horizon radius $r_+$ for different values of $\alpha$ and also discuss the first law of black hole thermodynamics for Bardeen black hoe in presence of thermal fluctuations. In section \ref{5}, we conclude our work  with final remarks. All the thermodynamic quantities are expressed in the units of $ \hslash= G = c=1 $.  
\section{thermodynamics of Bardeen regular black hole and its Geometrothermodynamic stability} \label{2}
The action for gravity coupled to some form of matter which could give rise to some kind of regular solution could be written as, \begin{equation}
S= \frac{1}{16 \pi}\int d^4x \sqrt{-g}(R-L),\end{equation} where $g$ is the determinant of the metric $g_{\mu \nu}$, $R$ represents curvature scalar. We now write down the general line element, $ ds^2= g_{\mu \nu}dx^\mu dx^\nu$ for spherically symmetric regular black hole as follows,
\begin{equation}
ds^2 = - f(r) dt^2 +\frac{dr^2}{f(r)} + r^2 d\theta^2 +r^2\sin^2\theta d\phi^2,
\end{equation}
with the metric function $f(r)$ taking the form as given below,
\begin{equation}
f(r)= 1- \frac{2m(r)}{r}.
\end{equation} The mass term $m(r)$ for Bardeen type regular back holes is of the form \cite{23},
\begin{equation}
m(r) = \frac{M}{(1+(\frac{a}{r})^q)^\frac{p}{q}},\label{m}.
\end{equation} 
To identify Bardeen and Hayward regular black holes, one specifies $p=3,q=2$ and $ p=q=3$ in mass term respectively. \\
As such we are left with the choice of setting $p=3,q=2$, thus our metric function takes the form of \cite{3},
\begin{equation} \label{metfn}
f(r) = 1- \frac{2 M} {r (1 + (\frac{a}{r})^2)^{3/2}}
\end{equation}
We first determine the event horizon radius by setting $f(r)$ = 0, hence we proceed as, 
\begin{equation} \label{horizon}
\left(a^2+r^2\right)^{3/2}-2 M r^2=0
\end{equation}
The positive real root of the equation \ref{horizon} will give us event horizon radius. We denote this positive real root by $r_+$. 
Now we discuss briefly the thermodynamics of Bardeen regular black hole. We will calculate thermodynamic quantities like entropy, free energy, internal energy, pressure, enthalpy and Gibbs energy. We first take the Hawking temperature ($T_H$) for Bardeen regular black hole from \cite{30}. Also in\cite{34a}, there is an elegant discussion of temperature of general class of Bardeen type regular black holes. Thus for our case, using standard definition \cite{nn}, $$ T_H = \frac{1}{4 \pi} \frac{df(r)}{dr}\bigg|_{r=r_+}$$,
\begin{equation} \label{temp}
T_H= \frac{M \left(r_+^3-2 a^2 r_+\right)}{2 \pi  \left(a^2+r_+^2\right)^{5/2}}.
\end{equation}
Here, $r_+$ reflects the greater among the two roots of the equation $f(r)=0$. If one evaluates the limit of $a
\rightarrow 0$, one should obviously get temperature for Schwarzchild black hole. Now, we  turn to calculate entropy of the Bardeen regular black hole using Bekenstein's are law \cite{7}. The expression for entropy can then be written as \cite{qqq}, 
\begin{equation}
S_0=\pi  r_+^2. \label{en}
\end{equation} 
We now calculate the internal energy for the Bardeen regular black hole. In fact, internal energy $E$ follows   the first law of black hole thermodynamics to arrive at the formula $E = \int T_HdS_0$. For a given temperature (\ref{temp}) and entropy (\ref{en}),  this reads,
\begin{equation} \label{eb}
E =	\frac{1}{2} \bigg [r_+-3 a \tan ^{-1}\bigg (\frac{r_+}{a}\bigg )\bigg ].
\end{equation}
Here we  set quantity $M = 1$ in order to make calculations more elegant. This choice must be in par with the condition of existence of horizon at this value of $M$.

Moreover, we  compute  an important thermodynamical quantity known  as Helmholtz free energy or free energy which is a measure of  useful work obtainable from a closed thermodynamic system. 
The standard definition of free energy $F$ is given by $F=-\int S_0 dT_H$.   For above values of $T_H$ and entropy $S_0$, the Helmholtz free energy is calculated as,
\begin{equation} \label{free}
F =\frac{1}{4} \bigg [r_+ \bigg (-\frac{3 a^2}{a^2+r_+^2}-1\bigg )+6 a \tan ^{-1}\bigg (\frac{r_+}{a}\bigg )\bigg ].
\end{equation}
  In non extended space one can easily derive pressure for any black hole thermodynamic system by using the definition \cite{bph}, $P= \frac{-TS}{2}$, which alternatively is equal to $P=-\frac{dF}{dV}$. Where $F$ is free energy and $V$ is thermodynamic volume. The pressure of Bardeen black hole thus becomes,
\begin{equation} \label{pressure}
P = \frac{2 a^4+7 a^2 r_+^2-r_+^4}{16 \pi  r_+^2 \left(a^2+r_+^2\right)^2}.
\end{equation}
 The enthalpy for a thermodynamical system is an important state function which gives an idea about the change in energy of system. The enthalpy  also provides an idea about the equilibrium conditions for the system. The enthalpy, in the context of black hole thermodynamics, gained prominence due to Kastor etal.\cite{kastor} when thermodynamic variables were included in first-law of black hole thermodynamics along with the cosmological constant. In this regard, it is argued  that the mass $M$ of an AdS black hole plays the role of enthalpy of classical thermodynamics.
Therefore, enthalpy for the system of black hole can be calculated from the formula $H=E+PV$. This for our case yields,
\begin{equation} \label{enthalpy}
H =	   \frac{8 a^4 r_++19 a^2 r_+^3-18 a \left(a^2+r_+^2\right)^2 \tan ^{-1}\left(\frac{r_+}{a}\right)+5 r_+^5}{12 \left(a^2+r_+^2\right)^2}.
\end{equation}
For the black holes to have a static boundary with fixed temperature, we need fixed pressure and temperature. In such  scenario, the relevant thermodynamic potential to be used is  the Gibbs potential  or Gibbs free energy.
Since the expressions for  Helmholtz Free energy, pressure and volume are known, so it is matter of 
calculation to derive Gibbs free energy  by  the formula $G=F+PV$. Here, we find that,
\begin{equation} \label{gibbs}
G = \frac{9 a \left(a^2+r_+^2\right)^2 \tan ^{-1}\left(\frac{r_+}{a}\right)-r_+ \left(5 a^4+4 a^2 r_+^2+2 r_+^4\right)}{6 \left(a^2+r_+^2\right)^2}.
\end{equation} 
Our next thermodynamical quantity of interest is
 heat capacity of Bardeen regular black holes. Heat capacity is important because this gives 
 an idea about the size of heat bath inside the thermodynamical system. In fact, heat capacity helps us to identify the regions of stability. 
The heat capacity at constant volume is defined by, 
\begin{eqnarray}
C = T_H\frac{\partial S}{\partial T_H} = T_H \frac{\partial S}{\partial r_+}\bigg (\frac{\partial T_H}{\partial r_+}\bigg )^{-1}.
\end{eqnarray}
This leads to the heat capacity for Bardeen black holes as,
\begin{equation}
C = \frac{r_+ \left(r_+^3-2 a^2 r\right)}{\left(a^2+r_+^2\right)^{5/2} \bigg [\frac{3 r_+^2-2 a^2}{2 \pi  \left(a^2+r_+^2\right)^{5/2}}-\frac{5 r \left(r_+^3-2 a^2 r_+\right)}{2 \pi  \left(a^2+r_+^2\right)^{7/2}}\bigg ]}.
\end{equation}
This value of heat capacity  can be analyzed  with respect to event horizon radius in order to identify the stable/unstable regions. The regions with positive specific heat define the stable regions. We note that $C$ takes positive values for,
 \begin{equation} \label{rep}
 0.370754 r_+ < a < \frac{r_+}{2^{1/2}}.
\end{equation}
It is now obvious that the size of remnants which we would discuss in next section should lie in the regions constrained by equation \ref{rep} ( for particular value of $a$). \\
We would now discuss the Geometrothermodynamics (GTD) of Bardeen black hole to study its stability. Geometrothermodynamics \cite{g1,g2,g3} is the geometric approach to study thermodynamics of any system. The geometric approaches to study the thermodynamics could be widely found in literature. Some examples include Herman \cite{g4}, Mrugala \cite{g5}, Weinhold \cite{g6} and Ruppeiner \cite{g7}. In Geometrothermodynamics a metric is constructed from a some thermodynamic potential by carrying first and second order partial derivatives with respect to some known extensive variables \cite{g8}. The main motive to use Geometrothermodynamic approach is because the thermodynamic potential used here is generally Legendre invariant. To proceed we write the expression for mass of Bardeen regular black hole in terms of entropy as, 
\begin{equation}
M=\frac{\left(\pi  a^2+S\right)^{3/2}}{2 \sqrt{\pi } S}.
\end{equation} Now with $M$ written as $M(S,a)$, we write the GTD metric as,  
\begin{equation}
g_{GTD} = \left(a M_a+S M_S\right)\left[
\begin{array}{cc}
-M_{\text{SS}} & 0 \\
0 & M_{\text{aa}} \\
\end{array}
\right].
\end{equation} Here the second order derivative terms could be explicitly written as,
$$M_{\text{SS}}=\frac{8 \pi ^2 a^4+4 \pi  a^2 S-S^2}{8 S^3 \sqrt{\pi ^2 a^2+\pi  S}}$$ and,
$$M_{\text{aa}}=\frac{3 \sqrt{\pi } \left(2 \pi  a^2+S\right)}{2 S \sqrt{\pi  a^2+S}}.$$ 
The stability of concerned system is then simply given by the sign of $g_{GTD}$. Also we would like to emphasize that GTD is also a powerful technique to study phase transition of any black hole thermodynamic system.  
\section{Remnants of Bardeen black holes}\label{22}
\begin{figure}[htb]
	\begin{center}$
		\begin{array}{cc}
		\includegraphics[width=75 mm]{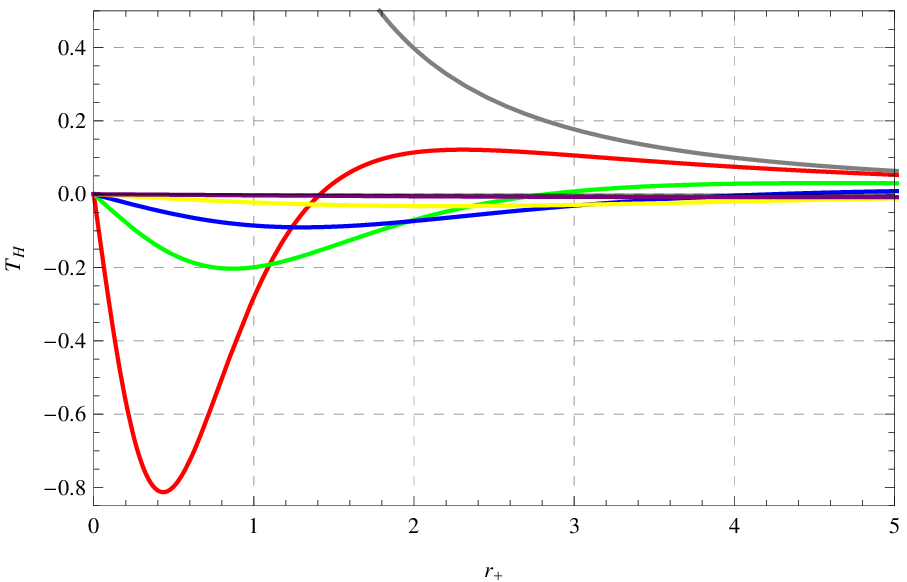}   \ \ \ \ & \includegraphics[width=75 mm]{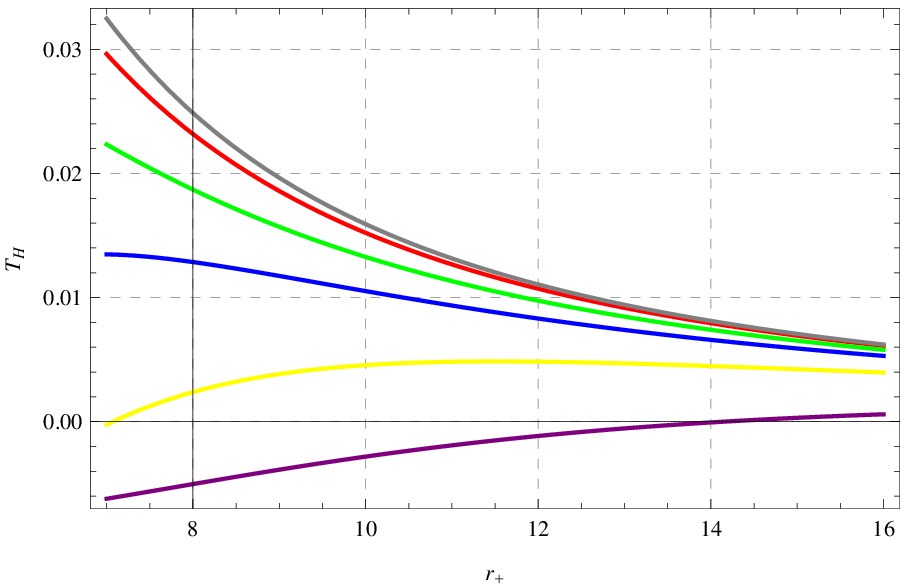} 
		\end{array}$
	\end{center}
	\caption{Left: Hawking temperature versus event horizon radius for Bardeen regular black hole for $M = 10$. Here, gray,  red, green, blue, yellow  and  purple lines correspond to $a = 0, 1, 2, 3, 5 $ and $10$, respectively. Here, it should  be noted that $a=0$ corresponds to the case od Schwarzchild black hole. Right: closeup of Left }
	\label{fig10} 
\end{figure}

\begin{figure}[htb]
	\begin{center}$
		\begin{array}{c}
		\includegraphics[width=80 mm]{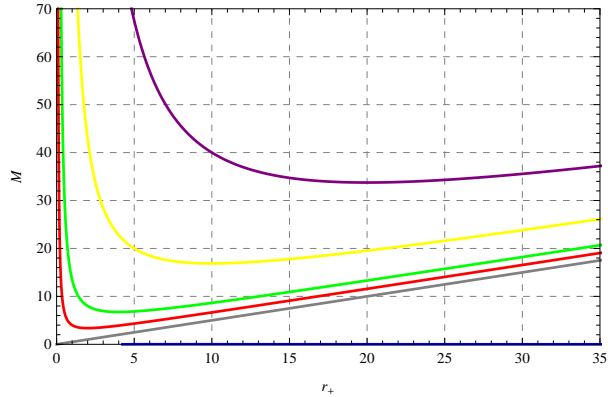}  
		\end{array}$
	\end{center}
	\caption{Mass  versus event horizon radius for Bardeen regular black hole. Here,   gray, red, green , blue, yellow  and purple lines  correspond to $ a =0, 1, 2, 3, 5$  and $10$, respectively.}
	\label{fig9}
\end{figure}

In this section, we  determine the remnant of Bardeen black hole. With the emission of Hawking radiation, the temperature eventually acquires a stage where the evaporation stops and the temperature gradually tends to zero value. At this stage,  black hole left with a stable remnant. There are various models based on quantum gravity approach which suggest that instead of  evaporating completely, the black hole turn out to become a remnant \cite{r0}. This remnant  exists only in the case when  we have finite value of event horizon radius at zero temperature. The radius of remnant can easily be  found by using the condition $T_H = 0$. The existence of remnant assures the existence of non-zero mass for any regular black hole (including Bardeen regular black hole), which could be either stable or long-lived \cite{r1}. The study of remnants becomes important in context that it could be thought as potential source of dark matter \cite{r2,r2a}. The black hole remnant could also prove helpful in resolving information loss paradox \cite{r3}. We now derive the expression for mass of Bardeen regular black hole as follows,
\begin{equation}
M= \frac{1}{2} r_+ \bigg [\frac{(a+r)^2}{r_+^2}\bigg ]^{3/2}.
\end{equation} 
To calculate Bardeen black hole remnant, we  use  $T_H = 0$. This results 
\begin{equation}
r_+=\sqrt{2}a.
\end{equation}
For this value of horizon radius $r_+$, the mass of remnant   $M_0$  is found to be 
\begin{equation}
M_0 = \frac{1}{4} \bigg [2 \sqrt{2}+3\bigg ]^{3/2} a.
\end{equation} 
We now plot temperature and mass of Bardeen regular black hole with respect to horizon radius $r_+$ in Figs. \ref{fig10} and \ref{fig9}, respectively. From the plot, it is evident that  for length parameter $a=0$ (Schwarzchild case)  temperature decreases along with the increase in the horizon radius. In fact,   temperature takes positive asymptotic value when horizon radius approaches to zero and black hole becomes colder along with larger horizon radius. 
In contrast, for non-zero values  of length parameter, the temperature of small black holes 
becomes negative valued and remains zero temperature for point like black holes.
As long as size of black hole 
increases after a critical value temperature becomes   positive valued.  The mass of Schwarzchild black hole increases 
with horizon radius. As long as $a$ takes non-zero values, the mass decreases first (abruptly) for small size black holes and then starts increasing with horizon radius.

\section{Corrected entropy for Bardeen black hole and stable remnants} \label{3}
\begin{figure}[htb]
	\begin{center}$
		\begin{array}{c }
			\includegraphics[width=80 mm]{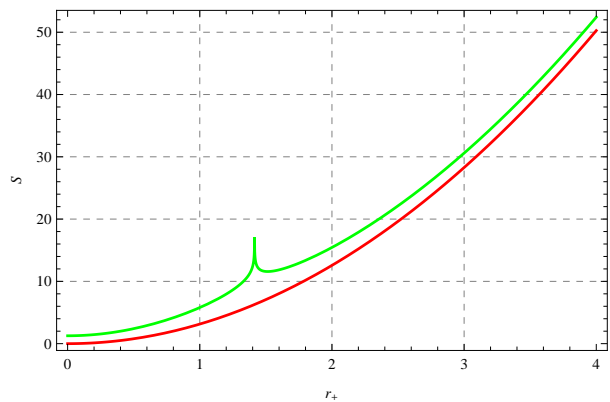}  
		\end{array}$
	\end{center}
	\caption{Entropy versus $r_+$ for $a=1$. Red and green  curves correspond  to $\alpha =0$  and $\alpha
		=1/2$, respectively.}
	\label{fig1}
\end{figure}

In this section, we  discuss the terms involved in leading-order corrected entropy for Bardeen regular black holes. To proceed, we start with partition function of Bardeen black hole. As such, we use density of states $\rho (E)$ and express partition function in terms of density of states as,
\begin{equation}
 Z(\beta) = \int_{0}^{\infty} dE \rho (E) e^{-\beta E}.
 \end{equation} 
 Here we have $\beta = \frac{1}{T_H}$ as Boltzmann constant is set unit. We now apply inverse Laplace transform to express density of states in terms of the energy of the underlying micro-states of the ensemble. Hence we proceed as, 
 \begin{equation} \label{a}
  \rho(E) = \frac{1}{2\pi i} \int_{\beta_0 - i\infty}^{\beta_0 - i\infty} d\beta Z(\beta)e^{\beta E} = \frac{1}{2\pi i} \int_{\beta_0 - i\infty}^{\beta_0 - i\infty} d\beta e^{S(\beta)},
 \end{equation} 
 where $S(\beta) = \ln Z + \beta E $ is the entropy of the system. Now using the method of steepest descent, we try to expand $S(\beta) $ around the equilibrium value of  $\beta$ (i.e. $\beta_0$). Thus we get,
 \begin{equation} \label{y}
 S(\beta) = S_0 +\frac{1}{2}(\beta - \beta_0)^2 \left. \frac{d^2S}{d\beta^2}\right|_{\beta = \beta_0} + \textnormal{higher-order terms}.
 \end{equation}
  We substitute the expression in \ref{y} in \ref{a}, which gives,
\begin{equation} \label{denf}
\rho(E) = \frac{e^{S_0}}{\sqrt{2\pi{\frac{d^2S}{d\beta^2}}}}.
\end{equation}
Taking logarithm of equation \ref{denf} we get the expression for entropy modified by thermal fluctuations as,
	\begin{equation} \label{w}
	S= \ln(\rho) = S_0 - \frac{1}{2} \ln \frac{d^2S}{d\beta^2} + \textnormal{other sub-leading terms.}
	\end{equation}
The second term in above equation is identified as correction term and in Ref. \cite{35}, it is estimated as,
\begin{equation}
	\frac{d^2S}{d\beta^2} = CT^2.
	\end{equation}
We note that (in regions with positive specific heat) the numerical    value of specific heat either coincides or remains proportional to the equilibrium value of canonical entropy.  Therefore,
\begin{equation} \label{q}
S= S_0 - \frac{1}{2} \ln S_0T^2.
\end{equation}
Without loss of generality, we use $\alpha$ in place of $\frac{1}{2}$  in the leading-order correction \ref{q}. This constant helps us to estimate and characterize the effect of thermal fluctuations in any thermodynamic system. Thus, the corrected entropy of the system due to thermal fluctuations then reads as,  
\begin{equation} \label{scorec}
S= S_0 - \alpha \ln S_0T^2.
\end{equation}
The logarithmic nature of leading-order correction  to the entropy  of   black hole system coincides with those in Refs. \cite{Kaul,carlip}. The  effect of thermal fluctuation on the thermodynamics is characterized by parameter $\alpha$. Plugging the values of equilibrium   entropy and Hawking temperature in Eq. (\ref{scorec}), the corrected   entropy  for Bardeen black hole takes following value:
\begin{equation}
S= \pi  r_+^2-\alpha  \ln \bigg [\frac{\left(r_+^2-2 a^2\right)^2}{16 \pi  \left(a^2+r_+^2\right)^2}\bigg].
\end{equation}
To analyze the   nature and effect of the  correction on entropy, we plot  figure \ref{fig1}.
From Fig. \ref{fig1}, we observe that the  change due to correction term tends to bring in stability in Bardeen black hole. The region where  the entropy curve  shows a kink towards positive Y-axis is a region of stability.  Remarkably, correction term insures the positive 
entropy even when black hole sinks towards becoming point like. In the limit $\alpha= 0$, one can recover equilibrium value (without any fluctuations). From the figure we can see that the region of stability is at $r_+ = \sqrt{2}a $ (for $a= 1$), which is exactly the black hole remnant radius for Bardeen black hole.  The thermal fluctuations have no effect for $r_+ < \sqrt{2}$, due to formation of a stable remnant.
\section{Leading order corrected thermodynamic equation of states and first law} \label{4} 
In this section, we calculate the effect of thermal fluctuation  on various thermodynamics quantities of Bardeen regular black hole. We first calculate the corrections in internal energy. The expression for uncorrected internal energy is given in  \ref{eb}. We calculate the corrected internal energy in the same way just by  using corrected entropy and obtain,
\begin{equation} \label{inc}
E_C =\frac{1}{4} \bigg [r_+ \left(2-\frac{3 \alpha }{\pi  r_+^2+\pi }\right)-\frac{3 (\alpha +2 \pi ) \tan ^{-1}(r_+)}{\pi }\bigg ].
\end{equation}
\begin{figure}[htb]
	\begin{center}$
		\begin{array}{c }
		\includegraphics[width=80 mm]{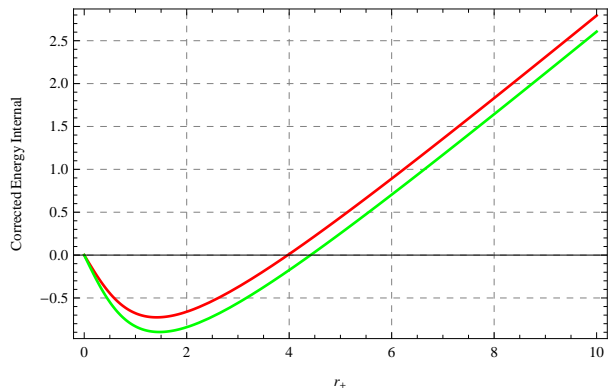}  
		\end{array}$
	\end{center}
	\caption{Internal energy versus $r_+$ for $a=1$. Red and green  curves correspond  to $\alpha =0$  and $\alpha
		=1/2$, respectively.}
	\label{fig2}
\end{figure}
We note from \ref{fig2} that internal energy of black hole reduces in the presence of thermal fluctuations. This decrease in energy is clear evidence of stability, which is reflected from first law of black hole thermodynamics \cite{barr}. By virtue of first law of black hole thermodynamics, black hole thermodynamic system could always be thought to be in equilibrium with black body radiations only at zero temperature. Now with some internal energy present in the black hole system, it would mean a non-zero temperature. Now at non zero temperature some radiation would cross the horizon and go into the black hole. Now in our case if the internal energy is decreasing, which means the temperature is approaching to zero. This suggests the attainment of equilibrium between black hole and black body radiations (Hawking radiations). This is a clear indication of stability in our thermodynamics system. 
Now after Internal energy  we move on to study the effect of thermal fluctuations on Helmholtz free energy or, more simply, free energy. The expression for uncorrected free energy of Bardeen regular black hole is derived in equation \ref{free}. The modified equation for free energy due to the thermal fluctuation can be estimated by using expression   $S$
 in place of  $S_0$ in the standard definition. This way the corrected free energy,  $F_C$ is 
 calculated by,
\begin{equation} \label{fc}
F_C =  \frac{2 r_+^2 \left(4 \pi  a^2-3 \alpha +\pi  r_+^2\right)+\alpha  \left(r_+^2-2 a^2\right) \ln \bigg [\frac{\left(r_+^2-2 a^2\right)^2}{16 \pi  \left(a^2+r_+^2\right)^2}\bigg ]}{8 \pi r_+ \left(a^2+r_+^2\right)}-\frac{3 \left(2 \pi  a^2+\alpha \right) }{4 \pi a}\tan ^{-1}\left(\frac{r_+}{a}\right).
\end{equation}
The detailed analysis of effect of thermal fluctuations on free energy of Bardeen regular black hole is depicted in Fig. \ref{fig3}. 
\begin{figure}[htb]
	\begin{center}$
		\begin{array}{c }
		\includegraphics[width=80 mm]{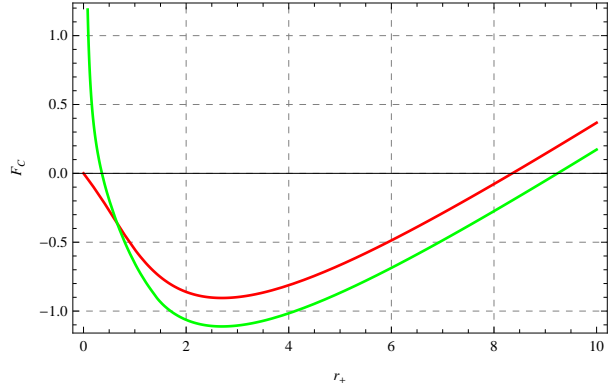}  
		\end{array}$
	\end{center}
	\caption{Helmholtz free energy versus $r_+$ for $a=1$. Red and green  curves correspond  to $\alpha =0$  and $\alpha
		=1/2$, respectively.}
	\label{fig3} 
\end{figure}
Here we see that the correction terms modify the behavior of the Helmholtz free energy for 
small sized black hole. In contrast to equilibrium case, the corrected Helmholtz free energy 
takes an asymptotically high value when horizon radius reduces to zero. 
 
\begin{figure}[htb]
	\begin{center}$
		\begin{array}{c}
		\includegraphics[width=80 mm]{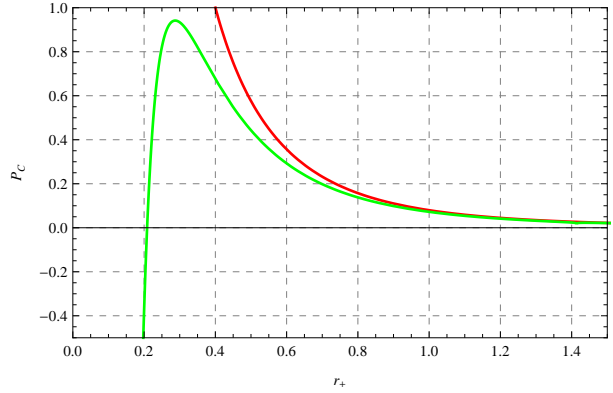}  
		\end{array}$
	\end{center}
	\caption{Pressure versus $r_+$  for $a=1$. Red and green  curves correspond  to $\alpha =0$  and $\alpha
		=1/2$, respectively.}
	\label{fig4}
\end{figure}

\begin{figure}[htb]
	\begin{center}$
		\begin{array}{c }
		\includegraphics[width=80 mm]{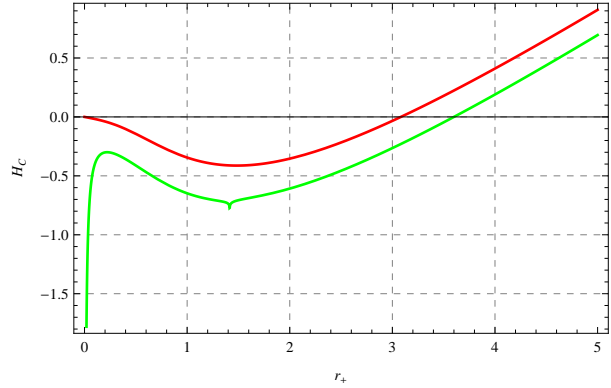}  
		\end{array}$
	\end{center}
	\caption{Enthalpy versus $r_+$ for $a=1$. Red and green  curves correspond  to $\alpha =0$  and $\alpha
		=1/2$, respectively.}
	\label{fig5}
\end{figure}

\begin{figure}[htb]
	\begin{center}$
		\begin{array}{c }
		\includegraphics[width=80 mm]{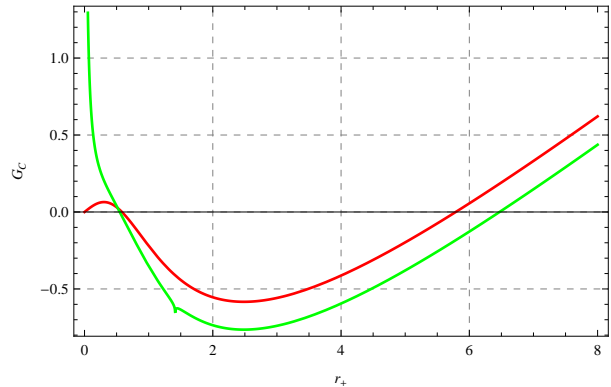}  
		\end{array}$
	\end{center}
	\caption{Gibbs free energy versus $r_+$ for $a=1$. Red and green  curves correspond  to $\alpha =0$  and $\alpha
		=1/2$, respectively.}
	\label{fig6}
\end{figure}

Furthermore,  we investigate the effect of fluctuations on pressure of Bardeen regular black hole. Corresponding to the corrected free energy $F_C$,   the corrected pressure $P_C$ is computed as,
\begin{equation}
P_C = \frac{2 r_+^2 \left[10 \pi  a^4+a^2 \left(6 \alpha +5 \pi  r_+^2\right)+\pi  r_+^4+6 \alpha  r_+^2\right]+\alpha  \left[2 a^4+7 a^2 r_+^2-r^4\right] \ln \left[\frac{\left(r_+^2-2 a^2\right)^2}{16 \pi  \left(a^2+r_+^2\right)^2}\right]}{32 \pi ^2 r_+^4 \left(a^2+r_+^2\right)^2}.
\end{equation}
In order to do comparative analysis, we plot $P_C$ with respect to $r_+$ in Fig. \ref{fig4}.
The effect of thermal fluctuation on the pressure becomes significant for the small sized black holes. Incidentally, the corrected pressure becomes negative when size of black hole sufficiently approaches zero radius. However, in absence of quantum correction the pressure is never negative in our case of Bardeen regular black hole (see the figure 6).
The enthalpy, which refers to total  heat   of a thermodynamic system, needs to be examined under thermal fluctuations.  The  corrected enthalpy  $H_C$ can be calculated following the same procedure as in the previous section. The only difference here is that we use corrected 
expressions of thermodynamical quantities. This gives,
\begin{eqnarray}
H_C &=& \frac{2 r_+^2 \left[10 \pi  a^4+a^2 \left(6 \alpha +5 \pi  r_+^2\right)+\pi  r_+^4+6 \alpha  r_+^2\right]+\alpha  \left[2 a^4+7 a^2 r_+^2-r_+^4\right] \ln \left[\frac{\left(r_+^2-2 a^2\right)^2}{16 \pi  \left(a^2+r_+^2\right)^2}\right]}{24 \pi  r_+ \left(a^2+r_+^2\right)^2} \nonumber\\
&- &
 \frac{3 \alpha  r_+}{4 \pi  \left(r_+^2+1\right)}-\frac{3}{4\pi } (\alpha +2 \pi ) \tan ^{-1}(r_+)+  \frac{ r_+}{2 }.
\end{eqnarray}
In order  to investigate the effect of thermal fluctuations on enthalpy of Bardeen black hole, we  plot the expression $H_C$ with respect to event horizon radius in Fig. \ref{fig5}.
From the figure, it is evident that the correction terms decrease the value of enthalpy of the system.
 There is a small kink present at $r_+ = \sqrt{2}$, which is the remnant radius for the given values of parameter $a$. This presence of small kink signifies the existence of stable remnant of Bardeen black hole. 
Finally, we discuss the effect of thermal fluctuations on  the Gibbs free energy. Following the
standard definition, we derive expression of corrected Gibbs free energy is calculated as,
\begin{eqnarray}
G_C &=& \frac{  r_+^2 \left[22 \pi  a^4+a^2 \left(20 \pi  r_+^2-3 \alpha \right)+4 \pi  r_+^4-3 \alpha  r_+^2\right]+\alpha  \left(-2 a^4+2 a^2 r_+^2+r_+^4\right) \ln \left[\frac{\left(r_+^2-2 a^2\right)^2}{16 \pi  \left(a^2+r_+^2\right)^2}\right] }{12 \pi   r_+ \left(a^2+r_+^2\right)^2}\nonumber \\
&-& \frac{ 3  \left(2 \pi  a^2+\alpha \right)  }{4 \pi  a   }\tan ^{-1}\left(\frac{r_+}{a}\right).
\end{eqnarray} 
The plot for $G_C$ versus event horizon radius is given in Fig. \ref{fig6}. The effect of correction terms on Gibbs energy of Bardeen black hole can clearly be depicted from figure.
 The decrease in Gibbs free energy is necessarily an indication of more stable thermodynamic system \cite{yan}. This indicates that thermal fluctuation make Bardeen black hole to decay into more stable black hole system.
  Hence presence of correction produces the desired results.
 We are concerned for the effect of thermal fluctuations near remnant radius, where in we can see from the figure the small kink is present. Also when size of black hole reduces to becoming point like, the Gibbs free energy becomes positive. In this region, the thermal correction becomes significant and reverts the behavior of Gibbs free energy. \\
 Now we would discuss the first law of black hole thermodynamics for Bardeen regular black holes in presence of thermal fluctuations. Using the formulation in \cite{l1,l2,l3}, we can write the expression for first law of thermodynamics as,
 \begin{equation} \label{fl}
 dM= TdS + PdV+ ......
 \end{equation}where $M,T,S, P, V$ denote mass, temperature, entropy,pressure and volume of black hole respectively. It is to be mentioned that the number of terms in R.H.S the first law of black hole thermodynamics could be more, if we include the charge and rotation degrees of freedom of black hole solution. We rewrite equation \ref{fl} as,
 \begin{equation} \label{xy}
 	X=	Y.
 \end{equation} One can easily see that in case of Bardeen regular black hole and in presence of thermal fluctuations $X \neq Y $, which means that $dM \neq TdS + PdV$ ( see Appendix). Hence we can say that in presence of thermal fluctuations, the first law of black hole thermodynamics is violated. However, we can find the various values of $r$ and $\alpha$ for which first law of black bole thermodynamics is satisfied. In that case our thermal fluctuations and first law would not be general in form but would become generic. For example to satisfy first law of black hole thermodynamics for Bardeen regular black holes in presence of thermal fluctuations at horizon radius $r=1$ and for $a=.5$, $\alpha$ must be equal to 1.42716. 
\section{Conclusion} \label{5}
We have explored the leading-order thermal fluctuations around equilibrium thermodynamics of Bardeen regular black holes. The thermal fluctuations for Hayward black holes can be easily found in literature \cite{34w}. We have identified the regions of stability for which the specific heat takes a positive value. We also recognized these regions by estimating the remnant radius for Bardeen black holes. We have found that for a particular value of horizon radius a stable remnant formed. The thermodynamic potentials accumulate their stable values at this particular horizon radius. We then tested this stable remnant employing thermal fluctuations. We anticipated that if the remnant is stable then the effect of thermal fluctuations should not distort the space-time geometries present inside the radius of the remnant. We observed that at this value of remnant radius the corrected entropy, which is a monotonically increasing function of event horizon radius, has a positive kink at $r_+= \sqrt{2}$. This is clear evidence of formation of a stable remnant at $r_+= \sqrt{2}$.   
This is also confirmed that the thermal fluctuations lead to an increase in entropy and hence, increases the stability of the system. Furthermore, we have studied the effect of thermal fluctuations on the internal energy of the Bardeen regular black hole. Here, we have observed that the internal energy of the black hole reduces in presence of thermal fluctuations. By using standard definition, the corrected Helmholtz free energy is also estimated under the effect of thermal fluctuations.   In contrast to the equilibrium value, the corrected free energy of the black hole takes (asymptotically) higher value when the horizon radius shrinks to zero. The corrected values of pressure, enthalpy, and Gibbs free energy are also computed. Then we also analyzed the first law of black hole thermodynamics for Bardeen regular black hole in presence of thermal fluctuations. We observed that first law is violated. We have found that the pressure becomes negative when the size of the black hole reduces.  The thermal fluctuations decrease the value of enthalpy of the system. Finally, we have remarked that the presence of thermal fluctuations increases the stability of black holes.
We thus conclude that unlike singular black holes, for which process of evaporation hence thermal radiation completely stops due to complete exhaustion of the mass of black hole when the radius of the event horizon approaches to zero,  the regular black holes still has some mass leftover when the temperature goes to zero. This mass is termed as the black hole remnant mass. This remnant is stable. The presence of thermal fluctuations further stabilize this remnant.   All the thermodynamic quantities tend to acquire stable values when the horizon of the black hole approaches to the remnant radius.

\appendix
\section{}
Here we would calculate and simplify the various terms involved in equation \ref{fl}, which is the first law of thermodynamics of Bardeen black hole in presence of thermal fluctuations. We treat the enthalpy of Bardeen black hole as mass and write the expression for mass ( using the corrected enthalpy ) as,
\begin{eqnarray}
M &= & \frac{2 r^2 \left[10 \pi  a^4+a^2 \left(6 \alpha +5 \pi  r^2\right)+\pi  r^4+6 \alpha  r^2\right]+\alpha  \left[2 a^4+7 a^2 r^2-r_+^4\right] \log \left[\frac{\left(r^2-2 a^2\right)^2}{16 \pi  \left(a^2+r^2\right)^2}\right]}{24 \pi  r\left(a^2+r^2\right)^2} \nonumber\\
&- &
\frac{3 \alpha  r_+}{4 \pi  \left(r_+^2+1\right)}-\frac{3}{4\pi } (\alpha +2 \pi ) \tan ^{-1}(r_+)+  \frac{ r_+}{2 }.
\end{eqnarray}Now differentiating w.r.t $r$, we get,
\begin{eqnarray} \label{xo}
dM&=&\bigg ( \frac{\left(10 a^6 r^2 \left(9 \alpha +\pi  \left(r^4+11 r^2+10\right)\right)-4 a^8 \left(\pi  \left(8 r^4+7 r^2-1\right)-9 \alpha \right)\right.)}{24 \pi  \left(r^3+r\right)^2 \left(r^2-2 a^2\right) \left(a^2+r^2\right)^3}\notag \\
&+& \frac{2 a^2 r^4 \left(3 \alpha  \left(r^4-r^2+1\right)+\pi  \left(2 r^4-5 r^2-7\right) r^2\right)+r^6 \left(\pi  r^2 \left(7 r^4-4 r^2-11\right)-6 \alpha  \left(r^4+5 r^2+1\right)\right)}{24 \pi  \left(r^3+r\right)^2 \left(r^2-2 a^2\right) \left(a^2+r^2\right)^3} \notag\\
&+& \frac{\left.3 a^4 r^2 \left(2 \alpha  \left(8 r^4+25 r^2+8\right)+\pi  \left(-11 r^6-4 r^4+7 r^2\right)\right)\right.}{24 \pi  \left(r^3+r\right)^2 \left(r^2-2 a^2\right) \left(a^2+r^2\right)^3}\notag \\
&+& \frac{\alpha  \left(r^2+1\right)^2 \left(4 a^8+4 a^6 r^2+45 a^4 r^4-26 a^2 r^6+r^8\right) \log \left(\frac{\left(r^2-2 a^2\right)^2}{16 \pi  \left(a^2+r^2\right)^2}\right)}{24 \pi  \left(r^3+r\right)^2 \left(r^2-2 a^2\right) \left(a^2+r^2\right)^3}\bigg ).
\end{eqnarray}Also we have, 
\begin{eqnarray}
T&=& -\frac{2 a^2-r^2}{4 \pi  a^2 r+4 \pi  r^3}.\\
dS&=& 2 \pi  r-\frac{8 \pi  \alpha  \left(a^2+r^2\right)^2 \left(\frac{r \left(r^2-2 a^2\right)}{4 \pi  \left(a^2+r^2\right)^2}-\frac{r \left(r^2-2 a^2\right)^2}{4 \pi  \left(a^2+r^2\right)^3}\right)}{\left(r^2-2 a^2\right)^2}.\\
V&=&\frac{4 \pi  r^3}{3}.  \\
dP&=& \frac{14 a^2 r-4 r^3}{16 \pi  r^2 \left(a^2+r^2\right)^2}-\frac{2 a^4+7 a^2 r^2-r^4}{4 \pi  r \left(a^2+r^2\right)^3}-\frac{2 a^4+7 a^2 r^2-r^4}{8 \pi  r^3 \left(a^2+r^2\right)^2}.
\end{eqnarray}We now calculate $Y$ of equation \ref{xy}, which is equal to 
\begin{eqnarray} \label{yo}
Y &=& \frac{4}{3} \pi  r^3 \left(\frac{14 a^2 r-4 r^3}{16 \pi  r^2 \left(a^2+r^2\right)^2}-\frac{2 a^4+7 a^2 r^2-r^4}{4 \pi  r \left(a^2+r^2\right)^3}-\frac{2 a^4+7 a^2 r^2-r^4}{8 \pi  r^3 \left(a^2+r^2\right)^2}\right)\notag \\&-&\frac{\left(2 a^2-r^2\right) \left(2 \pi  r-\frac{8 \pi  \alpha  \left(a^2+r^2\right)^2 \left(\frac{r \left(r^2-2 a^2\right)}{4 \pi  \left(a^2+r^2\right)^2}-\frac{r \left(r^2-2 a^2\right)^2}{4 \pi  \left(a^2+r^2\right)^3}\right)}{\left(r^2-2 a^2\right)^2}\right)}{4 \pi  a^2 r+4 \pi  r^3}
\end{eqnarray} Substituting \ref{xo} and \ref{yo} in equation \ref{xy}, we see that $ X\neq Y$. This means that in presence of thermal fluctuations, for Bardeen black hole $ dM \neq T dS + VdP$. Hence we can say that first law is violated. However, we can find the various values of $r$ and $\alpha$ for which first law of black bole thermodynamics is satisfied. As an example we would try to find the value of $\alpha$ for which first law of black hole thermodynamics for Bardeen black hole is satisfied. We proceed by dividing $dM$ by $ TdS + VdP$ to obtains expression for $W$, i.e
$$W= \frac{dM}{TdS + VdP}.$$ We now solve $W=1$ for $\alpha$ in terms of $a$ and $r$ and obtain,
\begin{equation}
\alpha = \frac{A}{B}
\end{equation}
\begin{equation}
A= 2 \pi  \left(r^4+r^2\right) \left(64 a^8 r^2+28 a^8+34 a^6 r^4-56 a^6 r^2+63 a^4 r^6+9 a^4 r^4-50 a^2 r^8-32 a^2 r^6+r^{10}+19 r^8\right)
\end{equation}
\begin{eqnarray}
B&=&  72 a^8 r^2-72 a^6 r^6+36 a^6 r^4-72 a^6 r^2+60 a^4 r^8+228 a^4 r^6+60 a^4 r^4+48 a^2 r^{10}+60 a^2 r^8+48 a^2 r^6\notag \\&+&r^{12} \log \left(\frac{\left(r^2-2 a^2\right)^2}{16 \pi  \left(a^2+r^2\right)^2}\right)-26 a^2 r^{10} \log \left(\frac{\left(r^2-2 a^2\right)^2}{16 \pi  \left(a^2+r^2\right)^2}\right)+2 r^{10} \log \left(\frac{\left(r^2-2 a^2\right)^2}{16 \pi  \left(a^2+r^2\right)^2}\right)\notag \\&-&52 a^2 r^8 \log \left(\frac{\left(r^2-2 a^2\right)^2}{16 \pi  \left(a^2+r^2\right)^2}\right)+r^8 \log \left(\frac{\left(r^2-2 a^2\right)^2}{16 \pi  \left(a^2+r^2\right)^2}\right)\notag \\&-&26 a^2 r^6 \log \left(\frac{\left(r^2-2 a^2\right)^2}{16 \pi  \left(a^2+r^2\right)^2}\right)+8 a^8 r^2 \log \left(\frac{\left(r^2-2 a^2\right)^2}{16 \pi  \left(a^2+r^2\right)^2}\right)+4 a^8 \log \left(\frac{\left(r^2-2 a^2\right)^2}{16 \pi  \left(a^2+r^2\right)^2}\right)\notag \\&+&4 a^8 r^4 \log \left(\frac{\left(r^2-2 a^2\right)^2}{16 \pi  \left(a^2+r^2\right)^2}\right)+4 a^6 r^2 \log \left(\frac{\left(r^2-2 a^2\right)^2}{16 \pi  \left(a^2+r^2\right)^2}\right)+4 a^6 r^6 \log \left(\frac{\left(r^2-2 a^2\right)^2}{16 \pi  \left(a^2+r^2\right)^2}\right)\notag \\&+&8 a^6 r^4 \log \left(\frac{\left(r^2-2 a^2\right)^2}{16 \pi  \left(a^2+r^2\right)^2}\right)+45 a^4 r^8 \log \left(\frac{\left(r^2-2 a^2\right)^2}{16 \pi  \left(a^2+r^2\right)^2}\right)+90 a^4 r^6 \log \left(\frac{\left(r^2-2 a^2\right)^2}{16 \pi  \left(a^2+r^2\right)^2}\right)\notag \\&+&45 a^4 r^4 \log \left(\frac{\left(r^2-2 a^2\right)^2}{16 \pi  \left(a^2+r^2\right)^2}\right)-12 r^{12}-60 r^{10}-12 r^8.
\end{eqnarray} Similarly we can solve $W=1$ for $r$ in terms of $\alpha$ and $a$.
\end{document}